\documentclass[conference]{IEEEtran}

\usepackage{csquotes}
\usepackage[natbib=true,backend=biber,style=ieee,sorting=none,giveninits=true]{biblatex}
\usepackage{hyperref}
\hypersetup{hidelinks}
\usepackage{microtype}
\usepackage[nameinlink,capitalize]{cleveref}
\usepackage{graphicx}
\usepackage{orcidlink}

\title{Threat-Oriented Digital Twinning for\\Security Evaluation of Autonomous Platforms
}

\author{%
\IEEEauthorblockN{Thomas J. Neubert, Laxima Niure Kandel\orcidlink{0000-0002-3933-1187} and Berker~Peköz\orcidlink{0000-0002-7572-3663}
}
\IEEEauthorblockA{Department of Electrical Engineering and Computer Science \\
Embry-Riddle Aeronautical University \\
Daytona Beach, FL, USA \\
E-mails: neubert2@my.erau.edu, \{Laxima.NiureKandel,Berker.Pekoz\}@erau.edu}
}

\begin{document}
\maketitle

\begin{abstract}
Open, unclassified research on secure autonomy is constrained by limited access to operational platforms, contested communications infrastructure, and representative adversarial test conditions. This paper presents a threat-oriented digital twinning methodology for cybersecurity evaluation of learning-enabled autonomous platforms. The approach is instantiated as an open-source, modular twin of a representative autonomy stack with separated sensing, autonomy, and supervisory-control functions; confidence-gated multi-modal perception; explicit command and telemetry trust boundaries; and runtime hold-safe behavior. The contribution is methodological: a reproducible design pattern that translates threat analysis into observable, controllable tests for spoofing, replay, malformed-input injection, degraded sensing, and adversarial ML stress. Although the implemented proxy is ground based, the architecture is intentionally framed around stack elements shared with UAV and space systems, including constrained onboard compute, intermittent or high-latency links, probabilistic perception, and mission-critical recovery behavior. The result is an implementable research scaffold for dependable and secure autonomy studies across UAV and space domains.
\end{abstract}


\section{Introduction}
Dependable autonomy now depends as much on the resilience of software, sensing, and communication paths as on the mechanical platform itself. Across UAV, loyal-wingman, and space systems, the same core problem recurs: learning-enabled autonomy introduces failure modes that are difficult to validate with static analysis alone, while access to fielded platforms is limited by cost, safety, and classification barriers. NIST guidance already frames cyber resilience and AI trustworthiness as system properties that must be engineered rather than assumed \cite{ross2021cyberresilient,tabassi2023airmf}. In parallel, digital twin literature has matured from a manufacturing concept to an assurance-oriented paradigm for high-value cyber-physical systems \cite{glaessgen2012digitaltwin,tao2019stateoftheart,voas2025digitaltwin}. The remaining gap is an open, unclassified autonomy twin that makes attack surfaces, trust boundaries, and degraded behaviors explicit.

Recent work has established three adjacent but still largely disconnected threads. First, security-oriented digital twins have been proposed for cyber-attack detection and anomaly analysis in cyber-physical systems, but these efforts are typically framed around industrial control or manufacturing environments rather than learning-enabled autonomous platforms \cite{balta2024dtattack,holmes2021cdt}. Second, runtime-assurance research has shown how monitored switching and safe fallback can preserve system safety when nominal autonomy becomes unreliable \cite{soter}. Third, UAV security research has studied communication spoofing and in-situ checking of airborne autonomy stacks, but usually at the level of individual links or firmware/runtime verification rather than as an end-to-end, open security evaluation environment \cite{huang2018combating,avis}. This paper addresses that gap with a threat-oriented autonomy twin that turns architectural trust assumptions into measurable assurance tests. Initial results show that authenticated session membership did not by itself guarantee telemetry provenance; higher-layer assurance still bounded the effect through degraded-mode and hold-safe transitions; and the same twin exposed additional assurance-relevant failure modes, including track starvation, stale-display persistence, and adversarial thermovisual perturbation, while supporting iterative hardening through retraining and threshold refinement.

The implementation is instantiated using a publicly documented robotic-combat-vehicle-like autonomy stack as it exposes the same reusable building blocks seen in UAV and space platforms: modular sensing, a central autonomy core, constrained onboard compute, supervisory command paths, and safety-critical degraded modes; and is therefore suitable for studying secure sensing and communications, ML robustness, runtime monitoring, spoofing and jamming resilience, and graceful recovery \cite{feickert2025rcv,dodd3000}.

The paper makes three claims. First, a useful security twin does not require perfect environmental realism; it requires faithful representation of the interfaces, states, and trust assumptions that govern system behavior under stress \cite{voas2025digitaltwin}. Second, threat-driven requirements produce a more defensible twin than simulation-first design because they force every major component to justify its observability, validation logic, and fail-safe behavior \cite{ross2021cyberresilient}. Third, an open-source implementation can still support meaningful security experimentation when it enforces explicit separation between perception, decision, and control, and when it records enough provenance and confidence metadata to evaluate why the system changed state.

\section{Threat-Oriented Requirements for an Open Twin}
The design begins with a simple systems-security assumption: in mission-critical autonomy, the most valuable twin is the one that preserves the attack surfaces most likely to produce unsafe or mission-degrading behavior. Three requirement classes dominate.

\emph{Perception integrity}: Learning-enabled autonomy should not consume raw detections as authoritative facts. It should expose confidence, support cross-modal agreement checks, and limit action when sensing becomes ambiguous. This requirement follows both AI risk-management guidance and the adversarial ML literature, which show that small perturbations or physically realizable manipulations can induce confident but unsafe misclassification \cite{tabassi2023airmf,goodfellow2015adversarial,eykholt2018physical}. The twin therefore treats perception outputs as candidate observations that must survive validation before influencing downstream state transitions.

\emph{Communication trust boundaries}: Contested links matter not only because they fail, but because they can fail selectively: stale commands may arrive late, replayed telemetry may remain well formed, and encrypted channels may still carry semantically malicious content. Threat modeling therefore focuses on spoofing, tampering, information disclosure, denial of service, and privilege escalation at the interface level, with ATT\&CK supplying a general adversary vocabulary and ATLAS complementing it for ML-enabled pipelines \cite{shostack2014threatmodeling,mitre_attack,mitre_atlas,huang2018combating}. The design therefore requires explicit separation of telemetry and command traffic, post-decryption validation, and auditable evidence of why records were accepted or dropped.

\emph{Runtime assurance and graceful degradation}: Autonomous systems require more than anomaly detection; they require bounded responses when trust in sensing, control, or timing decays. NIST cyber-resilience guidance emphasizes the ability to anticipate, withstand, recover, and adapt \cite{ross2021cyberresilient}. The design therefore requires explicit degraded-mode logic, bounded fallback behavior, subsystem liveness monitoring, and operator-visible supervisory control that keeps high-impact actions governable even under stress \cite{dodd3000,soter,avis}.

\section{Twin Architecture and Implementation Pattern}
The resulting architecture decomposes the platform across three isolated execution domains: a sensor/simulation host, an autonomy-core host, and a control/gateway host. This is a software analogue of the modular separation found in real autonomy programs, where sensing, decision-making, and supervisory control are coupled through interfaces but not collapsed into a monolith \cite{feickert2025rcv,skyborg2021}. Internal data exchange is implemented over ROS~2, which is attractive for open research because it preserves typed interfaces, modular nodes, and DDS-backed distributed communication while still supporting security extensions through SROS2 and DDS-Security \cite{ros2security,sros22022}. Mobility execution uses a standard open navigation stack, allowing the twin to represent end-to-end command-to-actuation flow without custom low-level motion code \cite{macenski2020marathon2}.

The perception path is intentionally multi-modal. RGB imagery feeds an ONNX-executed detector, while depth, LiDAR, and thermal streams provide corroborating geometric and physical context. The model is not granted direct authority. Instead, detections are converted into confidence-weighted tracks only after spatial consistency and cross-sensor checks succeed. This is a deliberate response to both ordinary sensor ambiguity and adversarial ML risk. An open twin cannot reproduce the full complexity of operational data, but it can reproduce the logic by which uncertain observations become trusted system state. To preserve reproducibility, the implementation uses controlled visual abstractions rather than sensitive operational datasets; the research value lies in the validation pipeline, not in benchmark-chasing object-recognition accuracy.

The communication path is split in two. Telemetry travels over a DTLS-protected channel from the autonomy core to the control gateway. Command traffic uses a separate MQTT/TLS supervisory path. That separation prevents the system from treating operator intent and autonomy-produced state as equivalent traffic classes, and it enables different validation policies for each. After DTLS termination, the gateway still checks length, schema, freshness window, source identity, and sequence monotonicity before data is published internally. This converts encrypted transport into an application-aware trust boundary consistent with zero-trust guidance that treats all communication paths as subject to explicit policy, authentication, and continuous validation \cite{rose2020zta}. Commands are likewise staged, serialized, and acknowledged explicitly rather than executed immediately from the operator interface.

\begin{figure*}
    \centering
    \includegraphics[width=\linewidth]{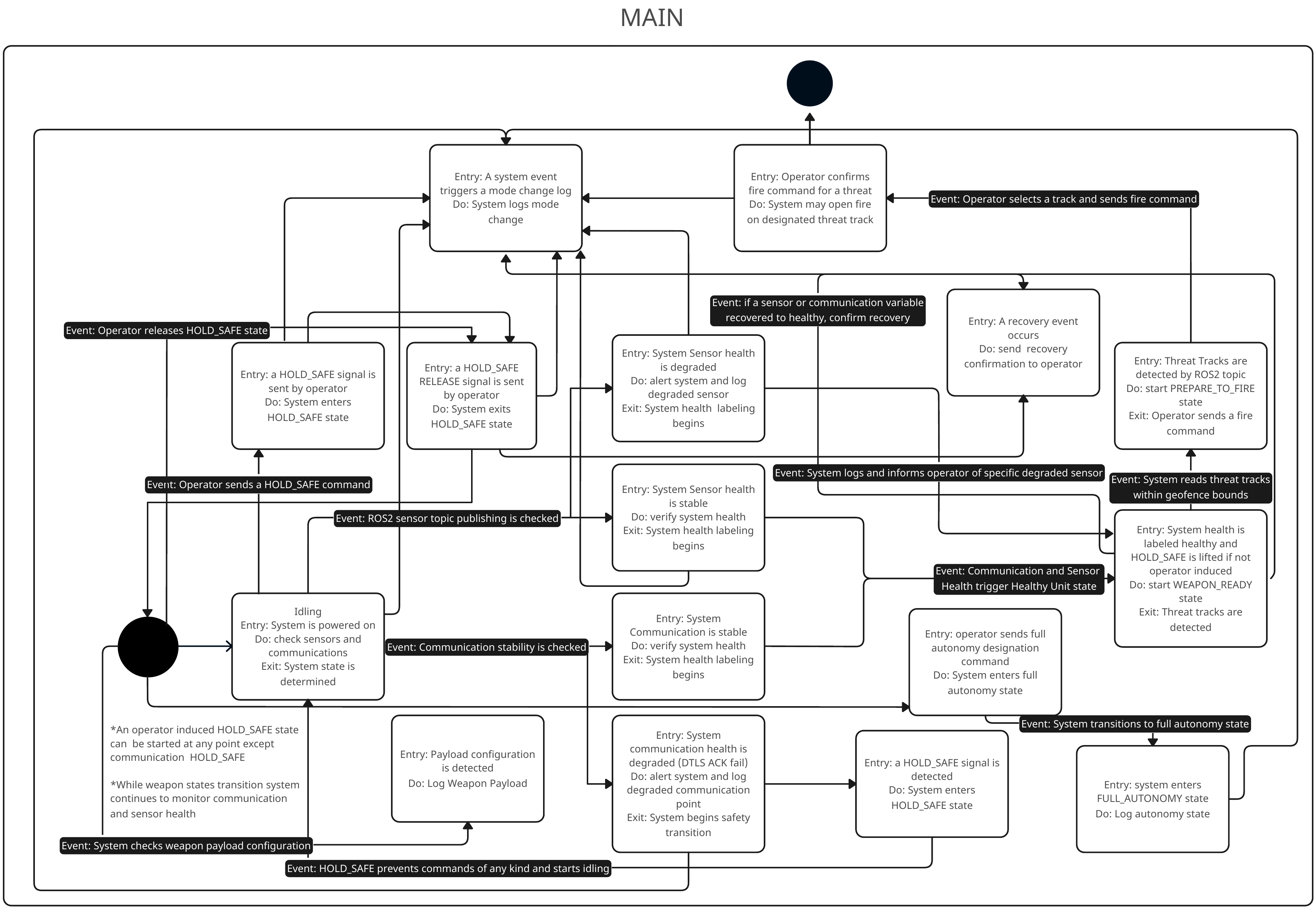}
    \caption{State transition diagram and decision logic implemented on the digital twin used for initial evaluation.}
    \label{fig:state}
\end{figure*}

The autonomy core itself is organized as a discrete state machine with health-aware transitions, a subset evaluated in this work presented in \cref{fig:state}. Candidate states include idle, ready, prepare-to-act, and restricted/hold-safe modes. Perception confidence, subsystem liveness, geofence status, and communication freshness influence these transitions. This design exposes the minimum machinery required to study how trust decays and how the system responds rather than attempting to replicate a classified policy engine. A small but useful addition is provenance metadata: track products carry mode and timing hints so downstream logic can distinguish likely live observations from synthetic or stale ones. This becomes valuable when studying replay, mode confusion, or inconsistent time bases.

\begin{figure*}
    \centering
    \includegraphics[width=\linewidth]{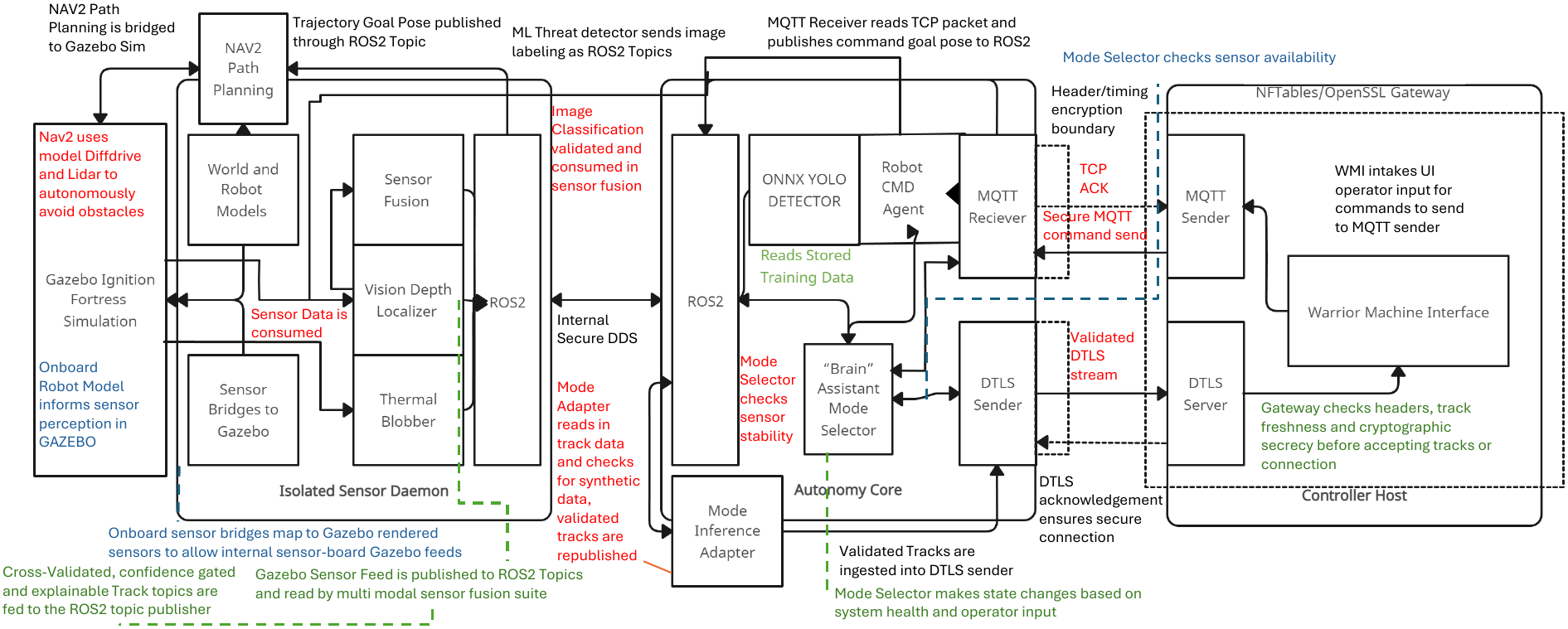}
    \caption{Threat-oriented digital twin architecture and trust boundaries.}
    \label{fig:arch}
\end{figure*}

Finally, the twin is instrumented for observability. Each major stage (perception, fusion, communication, operator interaction, and state transition) emits structured logs and machine-consumable summaries. This instrumentation turns the simulator into a security testbed by exposing whether a spoofed input was filtered at the gateway, whether confidence collapsed before action, and whether the platform entered hold-safe for the right reason. \Cref{fig:arch} summarizes the architecture and trust boundaries.
\section{Threat-to-Test Mapping and Initial Evaluation}
The initial evaluation focused on whether the architecture could expose cross-layer trust failures and bound their operational effect under representative communication and perception faults. It covered communication-path attacks, perception stress, and runtime-assurance behavior.
\subsection{Communication Path Attacks}
\begin{figure*}
    \centering
    \includegraphics[width=\linewidth]{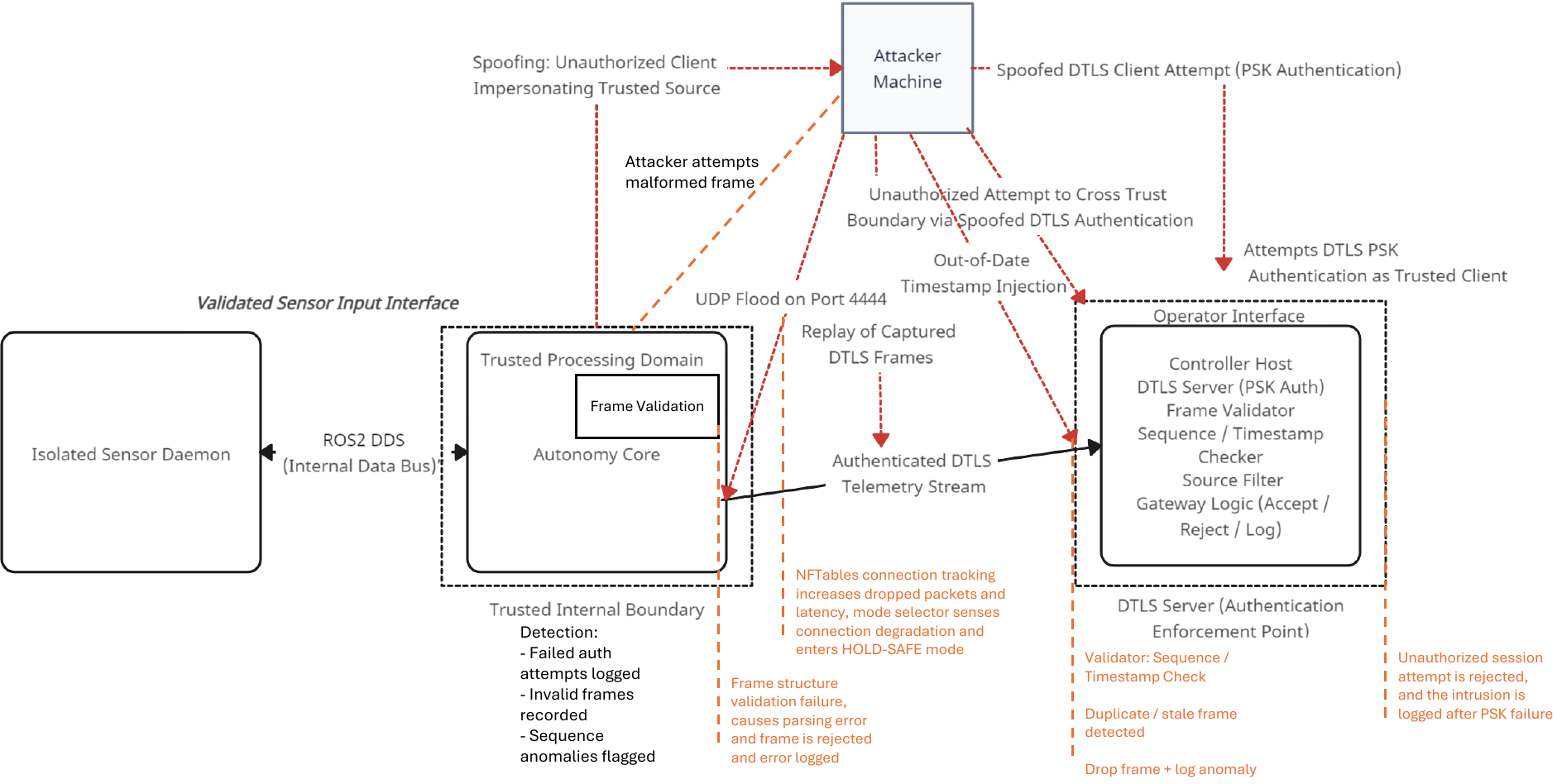}
    \caption{Threat-to-test mapping diagram for communication attacks.}
    \label{fig:threat}
\end{figure*}

A preliminary communication-path evaluation was conducted using two measures: application-level rejection of stale, malformed, or provenance-inconsistent telemetry; and safety-preserving containment, defined as dropping the record, downgrading trust, entering a degraded mode, or transitioning to \texttt{HOLD\_SAFE} before unsafe state propagation \cite{avis}. To instantiate these conditions without modifying autonomy or gateway logic, an adversarial relay was inserted on the DTLS telemetry path and used to introduce replay, delay, duplication, and packet-loss effects in a controlled and repeatable manner as shown in \cref{fig:threat}.

The more significant result arose in a teammate scenario. When a secondary teammate identity was admitted under the intended trust model, transport authentication was satisfied, but authenticated session membership did not by itself guarantee semantic provenance of the forwarded telemetry. Telemetry derived from another unit could therefore cross the cryptographic boundary when timing remained within the accepted freshness window. However, freshness, sequencing, and higher-layer consistency checks then drove the receiving unit into \texttt{HOLD\_SAFE} rather than allowing continued reliance on inconsistent teammate state. This is the central finding of the experiment: the threat-oriented twin exposed a cross-layer gap between transport identity and telemetry origin, and showed that its operational effect could still be bounded by containment logic above the cryptographic boundary.

As a negative control, a naive relay that re-originated traffic as a different apparent peer did not enter the established DTLS association and produced no new application-visible telemetry. This confirmed that the more interesting failure mode was not unauthenticated relay insertion, but authenticated provenance ambiguity that survived transport admission and was then contained at the application and state-machine layers \cite{balta2024dtattack,holmes2021cdt}.

\subsection{Perception Stress and Runtime Assurance}
The multi-modal pipeline also allows synthetic environmental changes, sensor dropout, and conflicting modality cues to be introduced in a controlled way. The aim is not to claim universal robustness, but to test whether confidence gating, cross-modal validation, and explainability metadata prevent a single manipulated observation from driving an unsafe state transition \cite{goodfellow2015adversarial,eykholt2018physical}. Freshness monitors, node liveness checks, geofence enforcement, and hold-safe logic then expose whether the platform fails conservatively when trust in sensing or timing decays.

\subsection{Evaluation Results}
Repeated subsystem-loss trials then quantified bounded degradation. Five thermal-processing-loss runs and five RGB-detector-loss runs each drove the platform into degraded operation and revoked \texttt{PREPARE\_TO\_FIRE} authority with no unsafe continuation. Mean degraded-transition latency was 511~ms for thermal loss and 957~ms for RGB-detector loss, with p95 latency below 1.7~s in both cases. Five localization/track-starvation trials also completed safely in degraded mode, although stale tracks occasionally persisted on the operator display. This exposed a layered assurance gap between display freshness and supervisory state, while still preventing unsafe engagement. Exploratory thermovisual perturbation further showed that baseline misclassification could be reduced through adversarially informed retraining; after retraining, manipulated cues were discarded without unsafe engagement, indicating that the remaining hardening need is plausibility-bounded thermal validation rather than additional mode logic.

\section{Cross-Domain Relevance to UAV and Space Systems}

The intended transferability of the twin is architectural rather than literal platform reuse: what generalizes across ground, UAV, and space systems is the assurance pattern of trust-bounded communications, freshness validation, confidence-weighted perception, supervisory safety control, observability of stale or degraded state, and graceful degraded-mode behavior. Those dependencies recur across autonomy stacks even when plant dynamics, sensor modalities, control laws, and timing regimes differ. For UAVs, the plant-facing layer would shift toward flight control, loiter, and return-to-base logic; for space systems, toward attitude control, payload management, and ground--space command mediation. The present results therefore show that an open twin can evaluate cross-layer assurance behavior at the interfaces most likely to survive platform changes.

\begin{figure*}
    \centering
    \includegraphics[width=\linewidth]{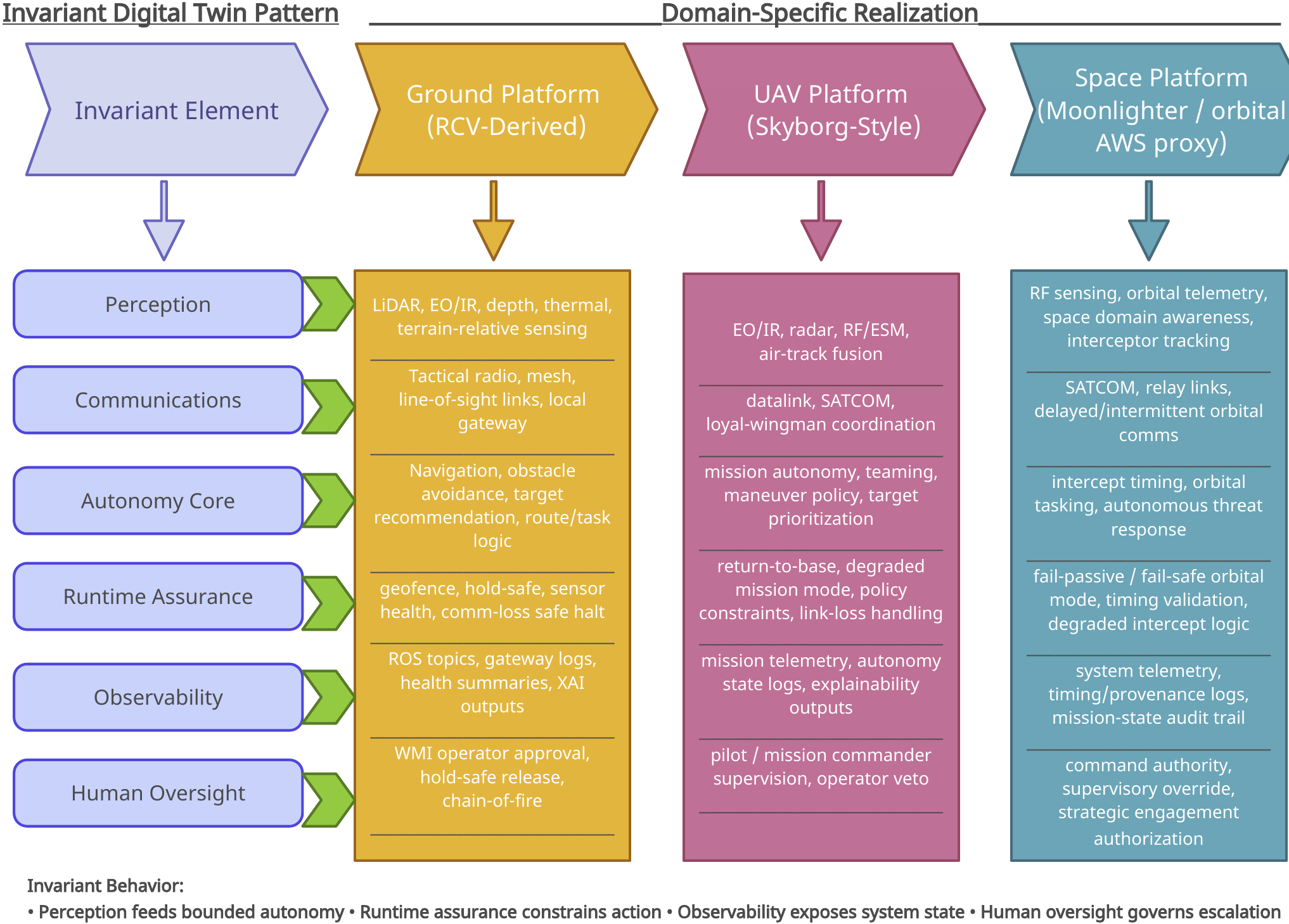}
    \caption{The universal autonomous-weapon-system digital twin pattern and its domain-specific realizations across ground, air, and space.}
    \label{fig:transfer}
\end{figure*}

\section{Conclusion}
This paper presented an open, threat-oriented digital twin for security evaluation of learning-enabled autonomous platforms and showed that the approach can expose trust-boundary weaknesses while preserving mission-safe behavior. The key empirical result is that transport admission did not by itself guarantee telemetry provenance in a teammate scenario, yet higher-layer assurance still bounded the operational effect through \texttt{HOLD\_SAFE} containment. Across repeated subsystem-loss trials, degraded transitions remained bounded, engagement authority was revoked consistently, and unsafe continuation was not observed. Taken together, these results show that a security-oriented twin can function as a practical pre-deployment assurance instrument by turning cross-layer trust assumptions, sensing failures, and supervisory controls into measurable resilience outcomes.

\printbibliography[title={References}]

\end{document}